# Estimating Traffic Conditions At Metropolitan Scale Using Traffic Flow Theory

**Weizi Li, Corresponding Author**
Department of Computer Science
University of North Carolina at Chapel Hill
201 S Columbia St, Chapel Hill, NC 27599
Email: weizili@cs.unc.edu

**Meilei Jiang**
Department of Statistics and Operations Research
University of North Carolina at Chapel Hill
Hanes Hall, 318 E Cameron Ave #3260, Chapel Hill, NC 27599
Email: jiangm@live.unc.edu

**Yaoyu Chen**
Department of Statistics and Operations Research
University of North Carolina at Chapel Hill
Hanes Hall, 318 E Cameron Ave #3260, Chapel Hill, NC 27599
Email: yaoyuchen@unc.edu

**Ming C. Lin**
Department of Computer Science
University of North Carolina at Chapel Hill
201 S Columbia St, Chapel Hill, NC 27599
Email: lin@cs.unc.edu

TRB paper number: 18-00974

Project website: gamma.cs.unc.edu/CityEst/

Word count: 1,178 words text +3 figures = 1,178 words

Submission Date: November 8, 2017

## INTRODUCTION

Traffic has become a major problem in metropolitan areas around the world. The extra cost due to traffic congestion and accidents are assessed over one trillion dollars worldwide. Therefore, understanding the complex interplay of road networks and travel conditions has been of a great interest in many contexts including improving simulation techniques *(1)*, analyzing urban infrastructure *(2)*, understanding human mobility *(3)*, and designing better routing strategies *(4)*. These applications highlight a need for developing a systematical framework that is capable of estimating traffic conditions using real-time sensor data.

We adopt GPS data for estimating citywide traffic conditions and use three processes to address the data features such as *low-sampling rate* and *spatial-temporal sparsity*: *map-matching (5,6,7,8)*, *travel-time inference (9,10,11,12,13)*, and *missing-value completion (6,14)*. While significant improvements have been achieved in these sub-areas, they are usually executed in *tandem*, which result in cascading errors and deteriorated estimations. Our framework conducts the estimation by first obtaining a coarse inference through a convex optimization program. Then, it refines the inferred vehicle paths and traffic conditions via *iteratively* performing *map-matching* and *travel-time inference*. Next, to handle the spatial sparsity, it conducts a nested optimization: the upper level aims to derive the optimal trip distributions among different areas in a road network while the lower level satisfies the constraints imposed by Wardrop Principles *(15,16)*. Finally, the framework addresses temporal sparsity using the Compressed-Sensing algorithm *(6)*.

We evaluate our framework using a real road network that consists of 5407 nodes and 1612 road segments, 34 heuristic network travel times corresponding to various congestion levels and times of a day, and over 10 million sampled GPS traces. The effectiveness of our approach has been compared to the state-of-the-art methods, namely Hunter et al. *(17)* and Rahmani et al. *(12)*, resulting in up to 96% relative improvements. To showcase our implementation, we conduct the field tests in Beijing and San Francisco using real-world GIS datasets, which contain 128701 nodes, 148899 road segments, and over 26 million GPS traces. The full report of this work can be found at gamma.cs.unc.edu/CityEst/ *(18)*.

## METHODOLOGY

Our goal is to estimate traffic conditions of a city-scale network. We explicitly address two challenges presented by GPS data: the low-sampling rate and the spatial-temporal sparsity through three steps: *coarse inference*, *iterative refinement*, and *nested estimation*. Traffic is commonly assumed to be quasi-static and has a weekly period *(17,19)*. Based on these observations, we divide an entire week into discrete time intervals and treat the traffic within each interval as static. In this section, we focus our discussion on the three steps over a single time interval. The temporal missing values over an entire traffic period are interpolated using the technique developed in *(6)*.

There are three key features of our work. First, for addressing the *low-sampling-rate* data, we use an iterative refinement rather than a sequential computation so that errors of the *map-matching* and *travel-time inference* processes are gradually reduced. Second, for addressing the spatial sparsity, we incorporate the sensing results generated from a large number of probe vehicles into the *traffic assignment* program so that we can compute travel times and flows of all road segments in a network. Third, our approach relies heavily on knowledge of transportation engineering, in which field the robust studies allow our modeling of traffic one step closer to the real-world traffic.

## EXPERIMENTS

In order to evaluate our approach, we use the road network from downtown San Francisco (obtained from openstreetmap.org) as the benchmark. The network contains 5407 nodes, 1612 road segments, and 296 TAZs (obtained from data.sfgov.org). We have also generated a set of heuristic network travel times using the System Optimal (SO) model and the Timestamp model based on the Cabspotting dataset (obtained from crawdad.org) as the ground truth.

We compare our work to Hunter et al. *(17)* and Rahmani et al. *(12)* using three measurements. The first metric is the performance gain of our technique over existing methods on travel times by considering all road segments of a network. The second metric is the error rate of the aggregate travel time of the entire network. The third metric regards the map-matching accuracy. The results are shown in Figure 1. Our method achieves as low as 8% error rate and as high as 96% relative improvement compared to the other two techniques.

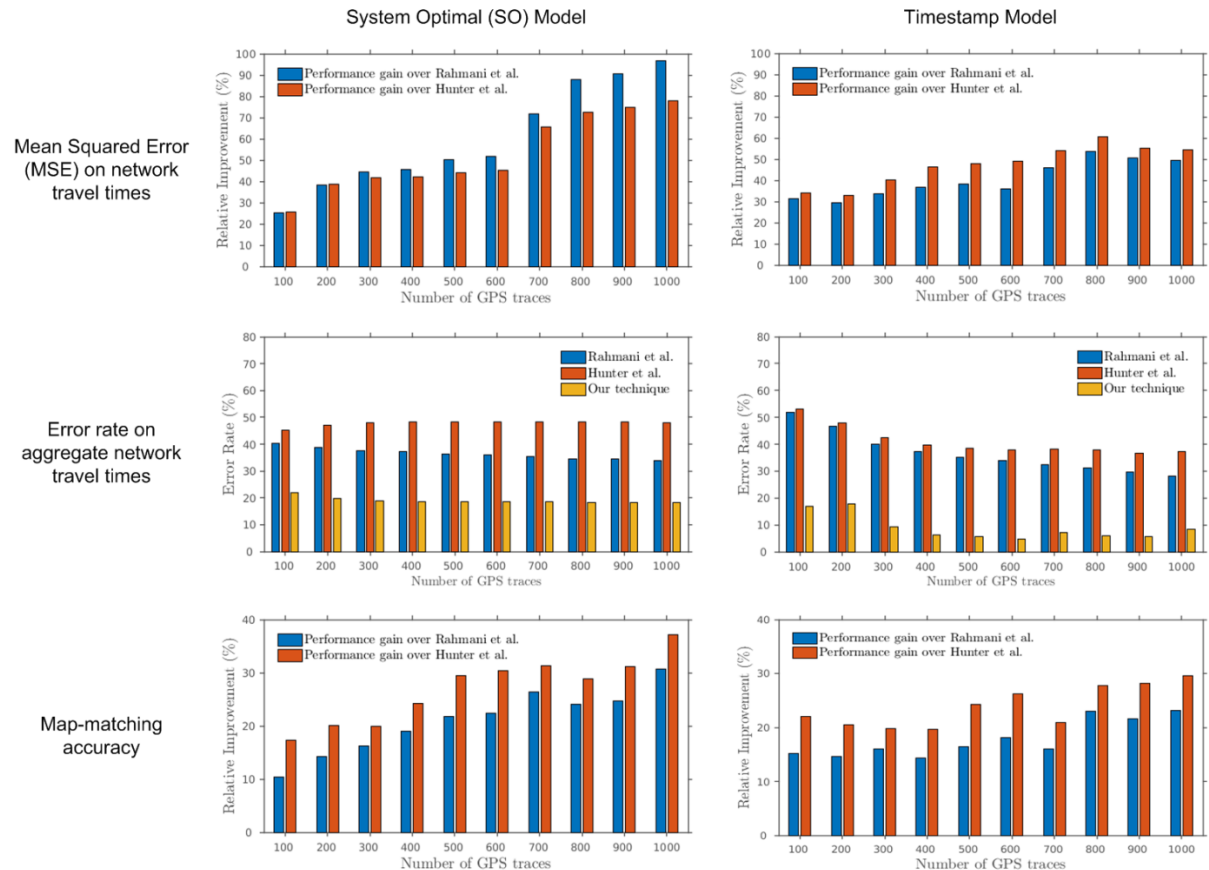

**FIGURE 1 From LEFT to RIGHT, LEFT diagrams show results via the System Optimal (SO) model, while RIGHT diagrams show results via the Timestamp model. TOP: The performance gain (%) of network travel times measured in Mean Squared Error (MSE). MIDDLE: The error rates (%) of all three methods on aggregate network travel times. BOTTOM: The performance gain (%) of map-matching accuracy measured using all sets of synthetic GPS traces. In summary, our technique achieves consistent improvements over the other two methods.**

## FIELD TESTS

We conduct field tests on two diverse cities in two continents, namely Beijing and San Francisco. The GPS datasets used in the field tests are from the Cabspotting project and T-drive project *(20)*, respectively.

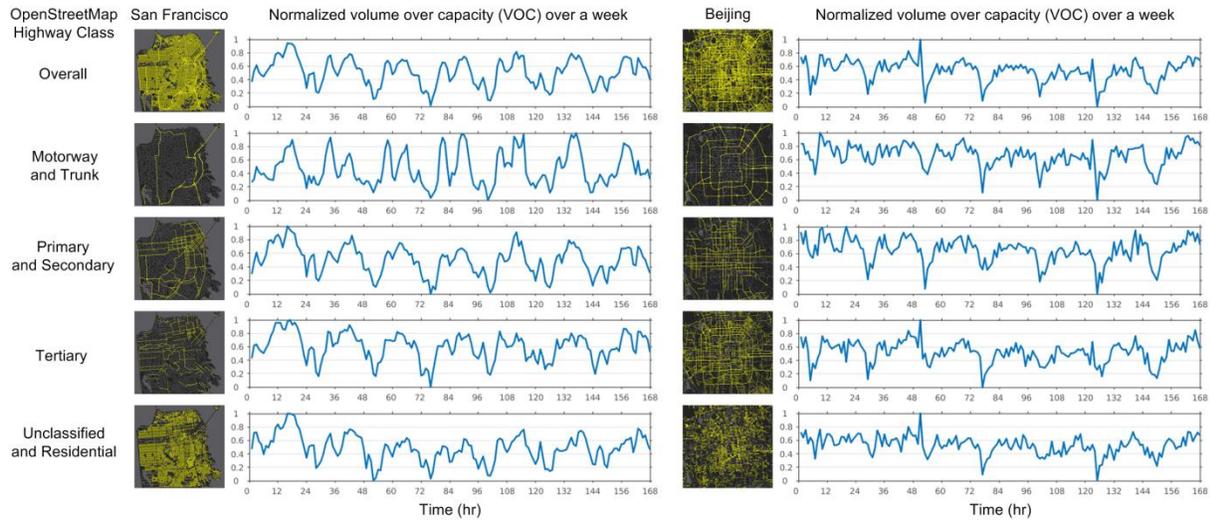

**FIGURE 2 The estimated traffic conditions measured in average volume over capacity (VOC) throughout the two cities, San Francisco and Beijing, for various types of roads. Our technique successfully recovers the periodic phenomena in all cases.**

The results shown in Figure 2 demonstrate the estimated traffic conditions measured in average volume over capacity (VOC) throughout the two cities. From the result, we can see that the recovered traffic patterns show clearly periodic phenomena over the course of a week – this feature is considered as one of the hallmarks of traffic *(19)* – for both overall roads and decomposed road types. In addition, in San Francisco, we observe saddle shapes corresponding to mid-day traffic relief over several days of a week. Such phenomena are more evident on major roads of San Francisco (i.e., *motorway and truck*), but not on the rest types of roads – on which the traffic patterns are similar indicating their similar usage as transportation infrastructure. In comparison, in Beijing, we don't observe such saddle shapes appearing in the middle of a day which suggests that congestion remains severe throughout the day time. Moreover, all types of roads of Beijing share similar traffic patterns indicating their similar usage in traffic.

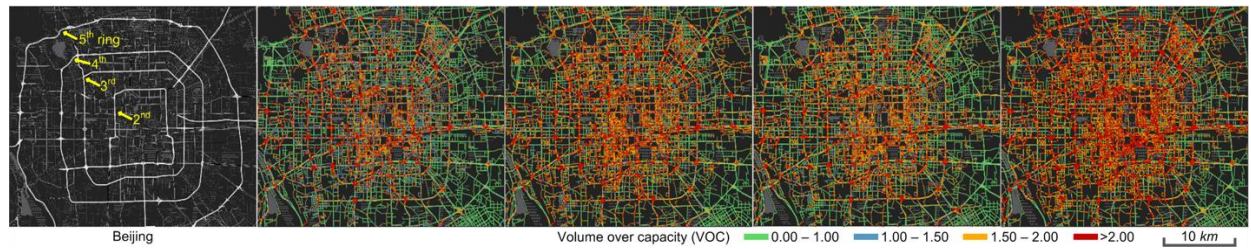

**FIGURE 3 Field tests in Beijing. Different colors represent different ranges of volume over capacity (VOC). Four time periods, namely Sunday 9AM, Tuesday 9AM, Thursday Noon,**

**and Friday 7PM are displayed as examples to illustrate weekend vs. weekday and morning vs. evening traffic.**

In Figure 3, we show detailed estimation results over four time intervals in Beijing: Sunday 9AM representing *weekend morning traffic*, Tuesday 9AM representing *weekday morning traffic*, Thursday Noon representing *weekday mid-day traffic*, and Friday 7PM representing *weekday evening traffic*. First, the Sunday morning's congestion tends to be the least severe and the Friday night's congestion tends to be the most severe. Second, the congestion situation of Thursday Noon is slightly better than Tuesday 9AM, especially considering the traffic between the 4th and the 5th ring roads, where more residential units are found than the region inside the 4th ring road.

## CONCLUSION AND FUTURE WORK

We have presented a novel framework for estimating urban traffic conditions using traffic flow theory and GPS traces. Our approach has been evaluated using a real road network resulting in consistent and notable improvements over state-of-the-art methods. In order to understand urban traffic patterns, two large-scale field tests were conducted in Beijing and San Francisco. Our estimated results can further enable traffic simulations and animations. Examples can be found in *(21)*.

There are several possible future directions. First of all, the coordination of probe vehicles in estimation can be explicitly taken into account. Second, with estimated traffic conditions, a *real-time* probabilistic mapping technique for GPS traces can be developed. Lastly, by fusing estimated results from historical data with accurate traffic simulations, it is possible to derive even more accurate forecasting of citywide traffic.

## ACKNOWLEDGMENTS

The authors would like to thank the National Science Foundation and US Army Research Office.